\documentclass[12pt]{article}
\usepackage[T1]{fontenc}

\usepackage[latin1]{inputenc}

\usepackage[english,french]{babel}

\usepackage{graphicx}

\usepackage{dcolumn}

\usepackage{amsmath,amssymb,amsfonts,amsthm}
\newcommand{\field}[1]{\mathbb{#1}}

\makeatletter

\@addtoreset{equation}{section}
\makeatother

\newtheorem{theorem}{Theorem}[]
\newtheorem{defin}{Definition}[]

\begin{document}


\selectlanguage{english}

\center{\Large{Quasiseparation of variables in the Schr\"{o}dinger
equation \\ with a magnetic field}} \vskip 1cm

{\setlength{\baselineskip}{0.5\baselineskip}
 \center{\small{F. Charest}} \center{\footnotesize{D\'epartement de 
math\'ematiques et de statistique}}
 \center{\footnotesize{Universit\'e de
Montr\'eal, C.P. 6128, succ. Centre-Ville}}
\center{\footnotesize{Montr\'eal, Qu\'ebec}}
\center{\footnotesize{H3C 3J7}} \center{\footnotesize{Canada}}
\center{\footnotesize{charest@dms.umontreal.ca}} \par}
{\setlength{\baselineskip}{0.5\baselineskip}
 \center{\small{C. Hudon}} \center{\footnotesize{D\'epartement de
physique}}
 \center{\footnotesize{Universit\'e de
Montr\'eal, C.P. 6128, succ. Centre-Ville}}
\center{\footnotesize{Montr\'eal, Qu\'ebec}}
\center{\footnotesize{H3C 3J7}} \center{\footnotesize{Canada}}
\center{\footnotesize{catherine.hudon@umontreal.ca}} \par}
\center{\small{and}} {\setlength{\baselineskip}{0.5\baselineskip}
 \center{\small{P. Winternitz}}
\center{\footnotesize{Centre de recherche math\'ematiques et
D\'epartement de math\'ematiques et de statistique}}
\center{\footnotesize{Universit\'e de Montr\'eal, C.P. 6128, succ.
Centre-Ville}} \center{\footnotesize{Montr\'eal, Qu\'ebec}}
\center{\footnotesize{H3C 3J7}} \center{\footnotesize{Canada}}
\center{\footnotesize{wintern@crm.umontreal.ca}}
\par}

\renewcommand{\baselinestretch}{1.0}.

\center{\large{\date{today}}} 

 \abstract{We consider
a two-dimensional integrable Hamiltonian system with a vector and
scalar potential in quantum mechanics. Contrary to the case of a
pure scalar potential, the existence of a second order integral of
motion does not guarantee the separation of variables in the
Schr\"{o}dinger equation. We introduce the concept of
"quasiseparation of variables" and show that in many cases it
allows us to reduce the calculation of the energy spectrum and
wave functions to linear algebra.}

\newpage

\section{Introduction}

A systematic search for integrable classical Hamiltonian systems
in magnetic fields was started quite some time ago \cite{Dorizzi,
McSween,Pucacco}.  A
Hamiltonian containing scalar and vector potentials was introduced
in a two-dimensional Euclidean space and additional integrals of
motion were constructed as linear or quadratic polynomials in the
momenta.  The same problem in quantum mechanics was investigated
quite recently \cite{Berube}.  Some new features have emerged in
 this study
of vector potentials that distinguish this case from that of
purely scalar ones.

1.  The existence of second order integrals of motion does not
imply the separation of variables in the Hamilton-Jacobi, or
Schr\"{o}dinger equation.

2.  Hamiltonian systems with second order integrals of motion in
classical and in quantum mechanics do not necessarily coincide
\cite{Berube}.

3.  In quantum mechanics an additional problem arises:  it is
necessary to choose a convenient gauge in which to write and solve
the Schr\"{o}dinger equation.

In the case of Hamiltonians with purely scalar potentials, quantum
and classical integrable systems may also differ, but only if the
integrals of motion are third, or higher order in the momenta
\cite{Hietarinta,Hietarinta2,Gravel,Gravel2}.  Moreover, third and
 higher order integrals of motion
are not related to the separation of variables, at least in
configuration space, in the case of purely scalar potentials
either \cite{Gravel,Gravel2}.

Variable separation in the presence of magnetic fields has also
been investigated \cite{Benenti,Zhalij,Zhdanov} and turns out to
 be quite rare.
Separable systems in this case only constitute a subset of the
quadratically integrable ones.

Natural questions that arise, specially in the context of quantum
mechanics, are the following.  What does one do with integrals of
motion that do not lead to the separation of variables?  How do
they help to integrate the Schr\"{o}dinger equation and find the
energy spectrum?

The purpose of this article is to provide at least a partial
answer to these questions.

The problem is formulated mathematically in Section 2.  Section 3
is devoted to systems with one first order operator $X$, commuting
with the Hamiltonian $H$.  In Section 4 the commuting operator $X$ is
assumed to be a second order operator of a specific ("cartesian")
type.  We introduce the concept of quasiseparation of variables
and identify all cases when quasiseparation occurs. The
Schr\"{o}dinger equation is solved for a specific scalar and
vector potential in Section 5. Section 6 is devoted to
conclusions.

\section{General setting}

The quantum Hamiltonian that we are considering is
\begin{eqnarray}
H=-\frac{\hbar ^2}{2}(\partial _x^2 + \partial _y^2)
-\frac{i\hbar}{2}(A\partial _x +\partial _xA +B\partial _y
+\partial _yB) + V,
 \label{Hamiltonien}
\end{eqnarray}
where A, B and V are functions of the coordinates x,y.  The
quantities of direct physical importance are the magnetic field
$\Omega$ and effective potential W:
\begin{eqnarray}
\Omega=A_y-B_x, \hspace{0.2cm} W=V-\frac{1}{2}(A^2+B^2).
\label{jauge}
\end{eqnarray}
These quantities are gauge invariant, i.e. unchanged by the
transformation
\begin{eqnarray}
\nonumber V\rightarrow \tilde{V} &=& V+(\textbf{A},\nabla
\phi) + \frac{1}{2}(\nabla \phi)^2,  \\
\textbf{A} \rightarrow \tilde{\textbf{A}} &=& \textbf{A}+\nabla
\phi, \hspace{0.2cm} \textbf{A}=(A,B), \label{gauge}
\end{eqnarray}
where $\phi$(x,y) is an arbitrary smooth function.

The classical equations of motion are
\begin{eqnarray}
\ddot{x}=\Omega\dot{y}-W_x, \hspace{0.2cm}
\ddot{y}=-\Omega\dot{x}-W_y \label{eqn_classique},
\end{eqnarray}
so all we need to know is $\Omega$ and W (the dots indicate time
derivatives).  In quantum mechanics we need the Hamiltonian
(\ref{Hamiltonien}).  Hence, given $\Omega$ and W, we must still
choose a gauge, i.e. the function $\phi$ in eq. (\ref{gauge}), and
calculate A, B and V.

The quantities $\Omega$ and W are obtained from the commutativity
condition
\begin{eqnarray}
[X,H]=0,
 \label{commutateur}
\end{eqnarray}
where X is the integral of motion, i.e. either a first, or a
second order linear differential operator.  The integral X is
obtained from the same condition.

The two equations to solve simultaneously, once a gauge is chosen,
are
\begin{eqnarray}
H\psi =E\psi, \hspace{0.2cm} X\psi=\lambda \psi, \label{systeme}
\end{eqnarray}
and the vector potential (i.e the gauge) should be chosen so as to
simplify this pair.

\section{First order integrals}

A first order integral of motion will have the form \cite{Berube}
\begin{eqnarray}
X=\alpha (L_3 +yA - xB) + \beta(P_1 + A) + \gamma( P_2+B) +m,
\label{x_1}
\end{eqnarray}
where $L_3$, $P_1$ and $P_2$ are the angular and linear momentum
operators, i.e.
\begin{eqnarray}
L_3=-i\hbar(y\partial _x -x\partial _y), \hspace{0.2cm}
P_1=-i\hbar\partial _x, \hspace{0.2cm} P_2=-i\hbar \partial _y,
\label{generateurs}
\end{eqnarray}
$\alpha$, $\beta$ and $\gamma$ are constants and the functions
m(x,y), $\Omega$ and W satisfy
\begin{eqnarray}
(\alpha x -\gamma )\Omega + m_x=0, \hspace{0.2cm} (\alpha y + \beta)\Omega +m_y=0,  \\
\label{cond1er}
 \nonumber (\alpha y +\beta )W_x +(-\alpha x +\gamma )W_y=0. \hspace{0.5cm}
\end{eqnarray}

Two inequivalent possibilities occur.  For $\alpha\neq$0 we can
put $\alpha$=1 and translate $\beta$ and $\gamma$ into
$\beta$=$\gamma$=0.  For $\alpha$=0, $\beta ^2$+$\gamma ^2\neq$0
we can rotate $\beta$ into $\beta$=0 and normalize $\gamma$ to
$\gamma$=1.  Let us now look at the two cases separately and solve
the Schr\"{o}dinger equation in each case.

a) $\alpha$=0, $\beta$=0, $\gamma$=1

We have \cite{Berube}
\begin{eqnarray}
m=m(x), \hspace{0.2cm} W=W(x), \hspace{0.2cm}
\Omega=\Omega(x)=\dot{m}(x). \label{condintcart}
\end{eqnarray}

The integral of motion (\ref{x_1}) in this case reduces to
\begin{eqnarray}
X=P_2+m(x)+B(x,y).
\end{eqnarray}
A convenient choice of gauge is
\begin{eqnarray}
B=-m(x), \hspace{0.2cm} A=0,
\end{eqnarray}
and from (\ref{jauge}) we obtain
\begin{eqnarray}
\Omega=\dot{m}(x), \hspace{0.2cm} V=W(x)+\frac{1}{2}m(x)^2.
\end{eqnarray}

The system (\ref{systeme}) reduces to
\begin{eqnarray}
\nonumber \{-\frac{\hbar^2}{2}(\partial_x^2+\partial_y^2)+i\hbar
m(x)\partial_y+W(x)+\frac{1}{2}m(x)^2\}\psi=E\psi,
\\ i\hbar\partial_y\psi=k\psi. \hspace{4cm}
\end{eqnarray}
With this choice of gauge we have the separation of variables,
i.e.
\begin{eqnarray}
\psi(x,y)=f_{Ek}(x)e^{-\frac{ik}{\hbar}y}, \label{sol_a}
\end{eqnarray}
\begin{eqnarray}
-\frac{\hbar^2}{2}\frac{d^2f}{dx^2}+[\frac{k^2}{2} +
km+W+\frac{1}{2}m^2-E]f=0. \label{edo1}
\end{eqnarray}

Equation (\ref{edo1}) is exactly solvable for many choices of m(x)
(i.e. for many choices of the magnetic field $\Omega$) and the
scalar potential V(x).  For instance, if we have
\begin{eqnarray}
m=\omega^2 x^2, \hspace{0.2cm} \Omega=2\omega^2 x, \hspace{0.2cm}
V(x)=0,
\end{eqnarray}
eq. (\ref{edo1}) has the solution
\begin{eqnarray}
f=e^{-\frac{\tau^2x^2}{2}}H_n(\tau x), \hspace{0.2cm}
\tau=\sqrt{\frac{\omega}{\hbar}}(2k)^{1/4}, \hspace{0.2cm}
E=\frac{k^2}{2}+(2n+1)\frac{\hbar \omega}{2}\sqrt{2k},
\end{eqnarray}
where $H_n$($\xi$) is a Hermite polynomial.  In this case f(x) is
regular and square integrable, however the solution (\ref{sol_a})
involves a plane wave in y.

b) $\alpha$=1, $\beta$=$\gamma$=0

In this case the determining equations (3.3) imply [2]
\begin{eqnarray}
m=m(r), \hspace{0.2cm} W=W(r), \hspace{0.2cm}
\Omega(r)=-\frac{\dot{m}}{r},
\end{eqnarray}
where we are using polar coordinates
\begin{eqnarray}
x=rcos\Theta, \hspace{0.2cm} y=rsin\Theta .
\end{eqnarray}
The operator (\ref{x_1}) in this case is
\begin{eqnarray}
X=L_3+yA-xB+m,
\end{eqnarray}
and a good choice of gauge is given by
\begin{eqnarray}
yA-xB+m=0,
\end{eqnarray}
leading to
\begin{eqnarray}
A(r,\Theta)=-\frac{m(r)}{r}sin\Theta, \hspace{0.2cm}
B(r,\Theta)=\frac{m(r)}{r}cos\Theta, \\
V(r,\Theta)=W(r)+\frac{m^2}{2r^2}. \hspace{3cm}
\end{eqnarray}

The solution of the system (\ref{systeme}) can be written in
separated forms as
\begin{eqnarray}
\psi(r,\Theta)=e^{-iM\Theta}\frac{1}{\sqrt{r}}R_{E,M}(r),
\end{eqnarray}
with R(r) satisfying
\begin{eqnarray}
-\frac{\hbar^2}{2}\frac{d^2R}{dr^2}+\{W(r) +
\frac{1}{r^2}[\frac{\hbar^2M^2}{2}+\frac{m(r)^2}{2} -\hbar m(r)M
-\frac{\hbar^2}{8}]\}R=ER. \label{edo2}
\end{eqnarray}

Again, this equation is exactly solvable in special cases, e.g.
\begin{eqnarray}
m(r)=\frac{\alpha}{r}, \hspace{0.2cm} W=-\frac{\alpha^2}{2r^4}.
\end{eqnarray}

The conclusion from this section is that first order integrability
in a magnetic field leads to a separation of variables either in
cartesian, or in polar coordinates.  To make this happen, a proper
choice of gauge is crucial.  Indeed, in a previous article [2] a
different choice of gauge was made, leading to R-separation,
rather than ordinary separation.  The separated ordinary
differential equations (\ref{edo1}) and (\ref{edo2}) both have the
form of one-dimensional Schr\"{o}dinger equations.  The magnetic
field and the effective potential combine together into an
x-dependent, or respectively r-dependent one-dimensional
"potential".

\section{Second order Cartesian integrability and the quasiseparation of variables}

Let us now consider a Hamiltonian of the form (\ref{Hamiltonien}),
admitting one second order integral of motion X.  It has been
shown that this operator X (or function in classical mechanics)
can be transformed into one of four standard forms [2,3]. They
were called Cartesian, polar, parabolic and elliptic, because in
the absence of a magnetic field, their existence leads to the
separation of variables in Cartesian, polar, parabolic or elliptic
coordinates, respectively \cite{Fris,Winternitz}.  The direct relation to the
separation of variables in the Schr\"{o}dinger, or Hamilton-Jacobi
equation does not hold in the presence of a magnetic field, but we
keep the terminology. In this article we restrict ourselves to the
Cartesian case.

The "Cartesian" integral of motion has the form \cite{Berube,Dorizzi}
\begin{eqnarray}
X=-\frac{\hbar^2}{2}\partial_x^2-i\hbar[(A+k_1)\partial_x+k_2\partial_y]-\frac{i\hbar}{2}
(A_x+k_{1x}+k_{2y})+\frac{1}{2}A^2 \label{x_2} \\
 +m+k_1A+k_2B.  \hspace{3.8cm}  \nonumber
\end{eqnarray}

All functions involved in the Hamiltonian (\ref{Hamiltonien}) and
the integral (\ref{x_2}) can be expressed in terms of two
functions of one variable each, $f=f(x)$ and $g=g(y)$, satisfying
\begin{eqnarray}
\ddot{f}=\alpha f^2+\beta f +\gamma, \hspace{0.2cm} g^{''}=-\alpha
g^2 +\delta g +\xi, \label{fg}
\end{eqnarray}
where $\alpha$, $\beta$, $\gamma$, $\delta$ and $\xi$ are real
constants.  We shall also use the first integrals of eq.
(\ref{fg}), namely
\begin{eqnarray}
\dot{f}^2=P_3(f), \hspace{0.8cm} g^{'2}=Q_3(g)  \hspace{4.0cm} \label{fg2}  \\
P_3(f)=\frac{2}{3}\alpha f^3+\beta f^2+2\gamma f+\sigma_1,  \hspace{0.8cm}
Q_3(g)=-\frac{2}{3}\alpha g^3+\delta g^2+2\xi g+\sigma_2 \nonumber
 \end{eqnarray}
where $\sigma_1$ and $\sigma_2$ are further real constants.  The
dots and primes are x and y derivatives, respectively.

Eq. (\ref{fg}) and (\ref{fg2}) can be solved in terms of elliptic
functions, or their degeneracies, if the polynomials $P_3(f)$, or
$Q_3(g)$ have multiple roots. In terms of $f(x)$ and $g(y)$ we have
\begin{eqnarray}
 \Omega  = \ddot{f}(x)+g^{''}(y), \hspace{4cm}\nonumber \\
  W = -\frac{\alpha}{3}(f-g)^3 -\frac{\beta+\delta}{2}(f-g)^2
  +(\xi-\gamma+\mu)(f-g), \hspace{0.5cm} \nonumber \\
 k_1  = -g^{'}(y), \hspace{0.2cm} k_2=-\dot{f}(x),\hspace{3cm}  \label{cond} \\
  m =
  -\frac{\alpha}{3}(g^3+2f^3-3gf^2)+\beta(fg-f^2)-\frac{\delta}{2}(f^2-g^2)
  \hspace{0.75cm} \nonumber \\
  -\gamma(2f-g)+\mu f +\xi g. \hspace{3.5cm} \nonumber
\end{eqnarray}
where $\mu$, figuring in the effective potential W and in m is an
additional constant.  The results (\ref{fg}) and (4.4) were
obtained in classical mechanics [3], but the classical and quantum
results coincide in the case of Cartesian integrability [2].

The vector potential (A,B) is yet to be chosen, but must satisfy
\begin{eqnarray}
A_y-B_x=\Omega=\ddot{f}(x)+g^{''}(y). \label{cond_jauge}
\end{eqnarray}
We shall in this section assume $\ddot{f}^2+g^{''2}\neq 0$.

Thus, we have a Hamiltonian $H$ and first integral X satisfying
(\ref{Hamiltonien}) and (\ref{x_2}), respectively, expressed in
terms of quantities satisfying eq. (\ref{cond}).  In general,
variables do not separate in eq. (\ref{systeme}) in any system of
coordinates.

Let us introduce the concept of "quasiseparation of variables". We
take a linear combination of the two equations (\ref{systeme}),
namely
\begin{eqnarray}
\{(H-E)+\phi(x,y)(X-\lambda)\}\Psi=0
\label{intermediaire}
\end{eqnarray}
where $\phi(x,y)$ is a function to be determined.  We wish to
choose the function $\phi(x,y)$ in such a manner that eq.
(\ref{intermediaire}) allows separation in Cartesian coordinates.
The solutions of eq. (\ref{intermediaire}) will then have the
separated form
\begin{eqnarray}
\Psi_{E\lambda}(x,y)=v_{E\lambda}(x)w_{E\lambda}(y).
 \label{separated}
\end{eqnarray}
The solution of the Schr\"{o}dinger equation will be a
superposition of separated solutions:
\begin{eqnarray}
\Psi_{E}(x,y)=\int
A_{E\lambda}v_{E\lambda}(x)w_{E\lambda}(y)d\mu(\lambda)
\label{superposition}
\end{eqnarray}
where $\mu(\lambda)$ is some measure to be chosen and
$A_{E\lambda}$ is independent of x and y.  For bound states the
integral in eq. (\ref{superposition}) will reduce to a sum.

On a more formal level we introduce the following definition.
\begin{defin} 
The commuting pair of operators $\{H,X\}$ allows the {\bf
quasiseparation of variables} in the system (\ref{systeme}) if
there exists a function $\phi(x,y)$ such that eq.
(\ref{intermediaire}) allows the separation of variables in the
sense of eq. (\ref{separated}).
\end{defin}

In this article we are considering the case when ($x$,$y$) are
Cartesian coordinates in a plane, but the concept of
quasiseparation is easily generalized to other coordinates and
other spaces.

Substituting (\ref{separated}) into (\ref{intermediaire}) we
obtain the equation
\begin{eqnarray}
-\frac{\hbar^2}{2}vw^{''}+\phi_1\ddot{v}w+\phi_2\dot{v}w+\phi_3vw^{'}+\phi_4vw=0
\label{sys1}
\end{eqnarray}
with
\begin{eqnarray}
 \phi_1=-\frac{\hbar^2}{2}(1+\phi), \hspace{0.25cm} \phi_2=-i\hbar A(1+\phi)+i\hbar \phi g^{'}, \hspace{0.25cm} \phi_3=-i\hbar (B-\phi\dot{f}), 
 \label{sys1_2}
\end{eqnarray}
\begin{eqnarray}
\phi_4=W+\frac{1}{2}A^2(1+\phi)+\frac{1}{2}B^2+\phi(m-g^{'}A-\dot{f}B)
\label{sys1_3} \\
\nonumber -\frac{i\hbar}{2}((1+\phi)A_x+B_y)-E-\lambda \phi.
\end{eqnarray}

The necessary and sufficient condition for variables to separate
in eq. (\ref{sys1}) is that we have
\begin{eqnarray}\label{sys2}
\frac{\phi_1}{\phi_4}=\frac{V_1(x)}{\tilde{V}(x)+\tilde{W}(y)}
\hspace{0.5cm}
\frac{\phi_2}{\phi_4}=\frac{V_2(x)}{\tilde{V}(x)+\tilde{W}(y)} \\
\nonumber
\frac{-\hbar^2}{2\phi_4}=\frac{W_1(y)}{\tilde{V}(x)+\tilde{W}(y)}
\hspace{0.5cm}
\frac{\phi_3}{\phi_4}=\frac{W_2(y)}{\tilde{V}(x)+\tilde{W}(y)}
\end{eqnarray}
where $V_i$, $W_i$, $\tilde{V}$ and $\tilde{W}$ are some
functions.

Let us consider the choice $\phi(x,y)=-1$ separately.

\subsection{$\phi(x,y)=-1$}

With this particular choice, eq. (\ref{sys1}) simplifies,
$V_2(x)$ reduces to a constant, and we find that

\begin{eqnarray}
B(x,y)= -\dot{f} + \frac{W_2(y)}{V_2} g^{'},\hspace{1cm}  and\hspace{1cm}  W_1(y)= \frac{\hbar}{2i}\frac{V_2}{g^{'}}. \label{B}
\end{eqnarray}

Condition (\ref{cond_jauge}) becomes
\begin{eqnarray}
A(x,y)= g^{'} + \tau(x), \label{A}
\end{eqnarray}
where $\tau(x)$ is arbitrary, and eqs. (\ref{sys2}) reduce to
\begin{eqnarray}
{\frac{g^{'}}{\phi_4}} = {- \frac{V_2}{i \hbar (\tilde{V}(x)+\tilde{W}(y))}}.
\end{eqnarray}

Thus
\begin{eqnarray}
\frac{\partial^2}{\partial x \partial y} (\frac{\phi_4}{g^{'}}) = 0
\end{eqnarray} 
is a necessary and sufficient condition for quasiseparation.
A straightforward calculation shows that this condition only
depends on g(y) and is equivalent to
\begin{eqnarray}
g^{''2} - g^{'} g^{'''} = 0.
\label{eq_g}
\end{eqnarray}
The general solution of eq. (\ref{eq_g}) is  $g(y)= C_1 e^{C_2 y} +
C_3$ and a two parameter class of particular solutions is $g(y)= g_1y
+ g_0$. It follows that
 $\phi(x,y)=-1$ only allows quasiseparation in the cases $\alpha=\delta=\xi=0$
 and  $\alpha=0$, $\delta>0$ (i.e. $g(y)= g_1 e^{\sqrt{\delta} y}+ g_2 e^{-\sqrt{\delta} y} + \frac{\xi}{\delta}$) with $g_1 g_2 =0$. 

\subsection{$\phi(x,y) \neq -1$}

Eliminating $\phi_4$ from the equations (\ref{sys2}) and using
relations (\ref{sys1_2}) and (\ref{sys1_3}) we obtain
\begin{eqnarray}
A(x,y)=\tau(x) + \frac{\phi(x,y)}{1+\phi(x,y)}g^{'}(y),
\hspace{0.25cm} B(x,y)=\eta(y)+\phi(x,y)\dot{f}(x), \label{sys3} \\
\nonumber
1+\phi(x,y)=\frac{\epsilon_1\sqrt{c_1-2kf(x)}}{\epsilon_2\sqrt{c_2+2kg(y)}}\sqrt{\frac{g^{'}(y)^2}{\dot{f}(x)^2}}, \hspace{1.5cm}
\end{eqnarray}
\begin{eqnarray}
V_1(x)=\epsilon_1 \sqrt{\frac{c_1-2kf(x)}{\dot{f}(x)^2}} ,
\hspace{0.5cm} W_1(y)=\epsilon_2 \sqrt{\frac{c_2+2kg(y)}{g^{'}(y)^2}}
 \label{sys3_2} \\
V_2(x)=\frac{2i}{\hbar}\tau(x)V_1(x), \hspace{0.5cm}
W_2(y)=\frac{2i}{\hbar}\eta(y)W_1(y). \nonumber
\end{eqnarray}
The functions $\tau(x)$ and $\eta(y)$ are arbitrary and can be
modified by gauge transformations (for instance they can be set
equal to zero).  The entries $c_1$, $c_2$, $k$, $\epsilon_1$ and
$\epsilon_2$ are constants with $\epsilon_1^2=\epsilon_2^2=1$.

The conditions (\ref{sys3}) on $A$, $B$ and $\phi$ are necessary
for separation of variables in eq. (\ref{sys1}).  There is a
further necessary condition that together with (\ref{sys3}) is
sufficient, namely
\begin{eqnarray}
W_1(y)\phi_4(x,y)=X(x)+Y(y),
 \label{sys4}
\end{eqnarray}
where $X(x)$ and $Y(y)$ are arbitrary functions.

In other words we must determine the conditions on the functions
$f(x)$, $g(y)$, $\tau(x)$, $\eta(y)$ and the constants $k$, $c_1$,
$c_2$, $\epsilon_1$, $\epsilon_2$ in such a manner that eq.
(\ref{sys4}) is satisfied.  All of the above quantities are real
and the square roots $\sqrt{c_1-2kf}$ and $\sqrt{c_2+2kg}$ must be
simultaneously real, or simultaneously imaginary.

To proceed further we use (\ref{sys1_3}), (\ref{sys3}) and
(\ref{sys3_2}) to obtain
\begin{eqnarray}
W_1(y)\phi_4(x,y)=R(x)+S(y)+T(x,y)
 \label{sys5}
\end{eqnarray}
\begin{multline}
R(x)=\\
-\frac{c_2\dot{f}^2(x)+(c_1-2kf(x))(2\lambda-2\sigma_1+\sigma_2-2\mu f(x)+\delta f^2(x)-\tau ^2(x)+i\hbar\dot{\tau}(x))}{2\epsilon_1\sqrt{c_1-2kf(x)}\sqrt{\dot{f}(x)^2}}
\label{sys5_2}
\end{multline}
\begin{multline}
S(y)=\\
-\frac{c_1g^{'2}(y)+(c_2+2kg(y))(2\epsilon-2\lambda+\sigma_1-2\sigma_2+2\mu g(y)+\beta g^2(y)-\eta^2(y)+i\hbar\eta^{'}(y))}{2\epsilon_2\sqrt{c_2+2kg(y)}\sqrt{g^{'}(y)^2}}
\label{sys5_3}
\end{multline}
and
\begin{eqnarray}
T(x,y)=\frac{F_1(x)g(y)}{\sqrt{c_1-2kf(x)}}+\frac{f(x)G_1(y)}{\sqrt{c_2+2kg(y)}}   \hspace{3.5cm} 
\nonumber \\
  \hspace{2.5cm}  +i\hbar[F_2(x)g^{'}(y) \sqrt{\frac{c_2+2kg(y)}{g^{'}(y)^2}}        +G_2(y) \dot{f}(x) \sqrt{\frac{c_1-2kf(x)}{\dot{f}(x)^2}}].
\label{sys5_4}
\end{eqnarray}
In (\ref{sys5_4}) we have
\begin{eqnarray}
F_1(x)= \frac{\epsilon_1[-k\dot{f}^2(x)+(c_1-2kf(x))\ddot{f}(x)] }{\sqrt{\dot{f}(x)^2}},         \\
\nonumber F_2(x)= \frac{\epsilon_2[k\dot{f}^2(x)+(c_1-2kf(x))\ddot{f}(x)]}{2(c_1-2kf(x))\dot{f}(x)},                              \\
\nonumber G_1(y)= \frac{\epsilon_2[kg^{'2}(y)+(c_2+2kg(y))g^{''}(y)]}{\sqrt{g^{'}(y)^2}},                              \\
\nonumber G_2(y)= \frac{\epsilon_1[kg^{'2}(y)-(c_2+2kg(y))g^{''}(y)]}{2(c_2+2kg(y))g^{'}(y)}.                                 
\label{sys5_5}
\end{eqnarray}
The separability condition (\ref{sys4}) is equivalent to the condition
\begin{eqnarray}
\frac{\partial^2T}{\partial x\partial y}=0.
\label{dp}
\end{eqnarray}

We can consider the real and imaginary parts of eq. (\ref{dp}) separately.  We obtain two conditions:
\begin{eqnarray}
\frac{1}{\dot{f}}\Big(\frac{F_1}{\sqrt{c_1-2kf}}\Big)^{.}=-\frac{1}{g^{'}}\Big(\frac{G_1}{\sqrt{c_2+2kg}}\Big)^{'}=N_1
\label{sys6}
\end{eqnarray}
\begin{eqnarray}
k \dot{F}_2 \sqrt{\frac{c_1-2kf}{\dot{f}^2}}=k G_2^{'}\sqrt{\frac{c_2+2kg}{g^{'2}}}=N_2,
\label{sys6_2}
\end{eqnarray}
where $N_1$ and $N_2$ are constants.

More explicitly, eq. (\ref{sys6}) and (\ref{sys6_2}) can be rewritten as
\begin{subequations}
\begin{equation}
\frac{-\epsilon_1[k^2\dot{f}^4+2k(c_1-2kf)\dot{f}^2\ddot{f}+(c_1-2kf)^2(\ddot{f}^2-\dot{f}\dddot{f})]}{(c_1-2kf)^{3/2}(\dot{f}^2)^{3/2}}=N_1 
\label{sys7a}
\end{equation}
\begin{equation}
\frac{\epsilon_2[k^2g^{'4}-2k(c_2+2kg)g^{'2}g^{''} +(c_2+2kg)^2(g^{''2}-g^{'}g^{'''})]}{(c_2+2kg)^{3/2}(g^{'2})^{3/2}}=N_1
\label{sys7b}
\end{equation} 
\begin{equation}
\frac{k\epsilon_2[2k^2\dot{f}^4+k(c_1-2kf)\dot{f}^2\ddot{f}-(c_1-2kf)^2(\ddot{f}^2 - \dot{f}\dddot{f})]}{2(c_1-2kf)^{3/2}\dot{f}^3}=N_2
\label{sys7c}
\end{equation}
\begin{equation} 
\frac{-k\epsilon_1[2k^2g^{'4}-k(c_2+2kg)g^{'2}g^{''}-(c_2+2kg)^2(g^{''2}-g^{'}g^{'''})]}{2(c_2+2kg)^{3/2}g^{'3}}=N_2
\label{sys7d} 
\end{equation}
\label{sys7}
\end{subequations}
We see that the functions $\tau(x)$ and $\eta(y)$ do not figure in (\ref{sys7}), and hence have no influence on the separation of variables.  The functions $f(x)$ and $g(y)$ depend on the constants $\alpha$, $\beta$, $\gamma$, $\delta$ and $\xi$ of eq. (\ref{fg}) and on a total of four further integration constants.  Our aim now is to find all values of these constants and of $k$, $c_1$ and $c_2$ for which eq. (\ref{sys7}) are satisfied.

Let us consider the cases $k=0$ and $k\neq0$ separately.

I) $k=0$

From (\ref{sys6_2}) we have $N_2=0$ and (\ref{sys7c},\ref{sys7d}) are satisfied identically.  Eq. (\ref{sys7a},\ref{sys7b}) simplify to
\begin{eqnarray}
N_1=\epsilon_1\sqrt{c_1}\frac{-\ddot{f}^2 + \dot{f}\ddot{f}}{(\dot{f}^2)^{3/2}}=\epsilon_2\sqrt{c_2}\frac{g^{''2}-g^{'}g^{'''}}{(g^{'2})^{3/2}}.
\label{sys8}
\end{eqnarray}
Using eq. (\ref{fg}) we obtain
\begin{eqnarray}
3\epsilon_1N_1\ddot{f}=2\alpha\sqrt{c_1} \sqrt{\dot{f}^2}, \hspace{0.5cm}  3\epsilon_2N_1g^{''}=2\alpha\sqrt{c_2} \sqrt{g^{'2}}.
\label{sys8_2}
\end{eqnarray}
Eq. (\ref{sys8_2}) are only compatible with (\ref{fg}) if we have $\alpha=N_1=0$.  The assumptions $c_1=0$ or $c_2=0$ lead to contradictions or $\phi(x,y)=-1$, so we are left with $\ddot{f}^2 - \dot{f}(\dddot{f})=0$, $g^{''2}-g^{'}(g^{'''})=0$ and hence, using (\ref{fg}) again we find that the only solutions for f(x) and g(y) are
\begin{eqnarray}
f(x)=f_1e^{\sqrt{\beta}x}+f_2e^{-\sqrt{\beta}x}-\frac{\gamma}{\beta} \nonumber \\
g(y)=g_1e^{\sqrt{\delta}y}+g_2e^{-\sqrt{\delta}y}-\frac{\xi}{\delta} \label{fg_exp} \\
\beta>0, \delta>0, f_1f_2=g_1g_2=0,  \nonumber
\end{eqnarray}
where $f_i$ and $g_i$ are constants, or one of the functions $f(x)$ or
$g(y)$ may be linear,
the other being as in (\ref{fg_exp}).

II) $k \neq 0$

We shall run through different possible solutions of eq. (\ref{fg}) and
determine which of them are compatible with eq. (\ref{sys7}).

1) $\alpha=\beta=\delta=0$, $\gamma\xi \neq 0 $.

We obtain
\begin{eqnarray}
f(x)= \frac{1}{2} \gamma x^2 \hspace{1.5cm} g(y)=\frac{1}{2} \xi y^2
\label{fg_quad}
\end{eqnarray}

Eq. (\ref{sys7}) and the condition $(c_2+2kg(y))(c_1-2kf(x))>0$ imply
\begin{eqnarray}
c_1=c_2=N_1=N_2=0,\hspace{1cm} \xi \gamma < 0
\end{eqnarray}

2) $\alpha=0$, \hspace{1cm} $\beta^2 + \delta^2 \neq 0$.

The functions f(x) and g(y) will be expressed in terms of exponentials,
trigonometric functions, or one of them may have the form (\ref{fg_quad}).
In none of these cases can eq. (\ref{sys7}) be satisfied.

3) $\alpha \neq 0 $.

In this case we are dealing with the two nonlinear equations (\ref{fg2}).
If the cubic polynomial on the right hand side has three distinct roots,
we obtain solutions in terms of elliptic functions. Otherwise the solutions
involve elementary functions.

Let us discuss the equation for f(x). We can assume with no loss of generality
that we have $\alpha>0$. Indeed, if we replace $f(x)\rightarrow -f(x)$,$\alpha \rightarrow -\alpha$,
$\beta \rightarrow \beta$,$\gamma \rightarrow -\gamma$,$\sigma_1 \rightarrow \sigma_1$ in eq. (\ref{fg2}) we
get the same equation. Hence we can change the sign of $\alpha$ from negative
to positive (if necessary). We write
\begin{eqnarray}
\dot{f}(x)^2=\frac{2}{3} \alpha (f(x)-f_1)(f(x)-f_2)(f(x)-f_3),
\hspace{0.5cm} \alpha>0
\end{eqnarray}
We require f(x) to be real, hence $\dot{f}(x)^2>0$. If all three roots are
real we order them as $f_1 \leq f_2 \leq f_3$.  Otherwise we consider $f_1\in \field{R}$
and $f_2=p+iq$, $f_3=p-iq$, $p,q \in \field{R}$, $q>0$. The possibilities are:

\begin{eqnarray}
a)\hspace{0.5cm} f_1 = f_2 = f_3 = -\frac{\beta}{2\alpha}, \hspace{0.7cm} f(x)=-\frac{\beta}
{2\alpha}+\frac{6}{\alpha (x-x_0)^2}  \hspace{0.7cm}
\end{eqnarray}

\begin{eqnarray}
b)\hspace{0.5cm} f_1 = f_2 < f_3 < f, \hspace{0.5cm} f(x)=f_1 +
\frac{f_3-f_1}{\sin^2{\omega(x-x_0)}}, \hspace{1.9cm} \\
 \nonumber f_1= \frac{-\beta + 2\sqrt{\beta^2 - 4\alpha \gamma}}{2 \alpha},
 \hspace{0.5cm} \omega^2 = \frac{\sqrt{\beta^2 - 4\alpha \gamma}}{4}, \hspace{0.5cm}
 (f_3-f_1)=\frac{6\omega^2}{\alpha}
\end{eqnarray}

\begin{eqnarray}
c)\hspace{0.5cm} f_1 \leq f \leq f_2 = f_3, \hspace{0.5cm} f(x)=f_1+(f_3-f_1)\tanh^2{\omega(x-x_0)},  \label{c} \\
  f_1= -\frac{\beta + 2\sqrt{\beta^2-4\alpha \gamma}}{2 \alpha}\hspace{0.5cm} \omega^2=\frac{\sqrt{\beta^2-4 \alpha \gamma}}{4} \hspace{0.5cm}
 (f_3-f_1)=\frac{6 \omega^2}{\alpha} \nonumber
\end{eqnarray}

\begin{eqnarray}
d)\hspace{0.5cm} f_1 < f_2 = f_3 \leq f, \hspace{0.5cm}
f(x)=f_1+\frac{f_3-f_1}{\tanh^2{\omega(x-x_0)}},  \hspace{1.4cm}
\end{eqnarray}
with $f_1$, $f_3$ and $\omega$ as in (\ref{c}).  
In cases $b$,$c$ and $d$ we assume $\beta^2-4\alpha\gamma>0$.

\begin{eqnarray}
e)\hspace{0.5cm} f_1 \leq f \leq f_2 < f_3, f_1 < f_2 \hspace{0.5cm}
f(x)=f_1+(f_2-f_1)sn^2(\omega(x-x_0),k),\label{f} \\
 \omega^2= \frac{\alpha}{6}(f_3-f_1), \hspace{0.5cm}
 k=\sqrt{\frac{f_2-f_1}{f_3-f_1}},  \hspace{2.5cm} \nonumber 
\end{eqnarray}
where $sn(\omega x,k)$ is a Jacobi elliptic function.

\begin{eqnarray}
f)\hspace{0.5cm} f_1 < f_2 < f_3 \leq f, \hspace{0.5cm} f(x)=f_1+\frac{f_3-f_1}
{sn^2(\omega(x-x_0),k)},  \hspace{2.5cm}
\end{eqnarray}
with k and $\omega$ as in (\ref{f}).

\begin{eqnarray}
g)\hspace{0.5cm} f_1 \in \field{R},\hspace{0.5cm} f_{2,3}=p \pm iq,\hspace{0.5cm} q>0,\hspace{0.5cm}
f(x)=f_1 + A\frac{1-cn(\rho x,k)}{1+cn(\rho x,k)},  \hspace{0.5cm} \\
\nonumber \rho=\sqrt{\frac{2\alpha}{3}A},\hspace{0.5cm} k^2=\frac{A-f_1+p}{2A},\hspace{0.5cm}
A^2=(p-f_1)^2+q^2  \hspace{1.0cm}
\end{eqnarray}

The solutions for $g(y)$ are similar and can be obtained from those for $f(x)$
by the substitutions
\begin{eqnarray}
f(x) \rightarrow -g(y),\alpha \rightarrow -\alpha,
\beta \rightarrow \delta,\gamma \rightarrow -\xi,\sigma_1 \rightarrow \sigma_2
\end{eqnarray}

The solutions for $f(x)$ and $g(y)$ can now be substituted into eq. (\ref{sys7})
in order to determine whether there exist constants $c_1$, $c_2$, k and $\epsilon_1$, $\epsilon_2$ for which the quantities $N_1$ and $N_2$ are indeed constant. It turns
out that if $f(x)$ or $g(y)$ are given by elliptic functions, eq. (\ref{sys7})
are never satisfied. However, if both polynomials in eq. (\ref{fg2}) have
multiple roots, eq. (\ref{sys7}) can always be satisfied. We give the results
for $f(x)$ and $g(y)$ in Tables 1 and 2, respectively. The calculations are
quite cumbersome and were performed using Mathematica.

From the tables, we see that any $f(x)$ from Table 1 can be combined with any
$g(y)$ from Table 2. The fact that $N_2$ must be the same in both tables
provides a relationship between $\epsilon_1$ and $\epsilon_2$.

More specifically, we have:
\begin{eqnarray}
\epsilon_1=\epsilon_2 \hspace{0.5cm} for \hspace{0.5cm}(F_1,G_3),(F_2,G_3),
(F_3,G_1),(F_3,G_2), \label{sol1} \\
(F_3,G_4),(F_4,G_3) \nonumber
\end{eqnarray}
\begin{eqnarray}
\epsilon_1=-\epsilon_2 \hspace{0.5cm} for \hspace{0.5cm}(F_1,G_1),(F_1,G_2),
(F_1,G_4),(F_2,G_1),(F_2,G_2), \label{sol2} \\
(F_2,G_4),(F_3,G_3),(F_4,G_1),(F_4,G_2),(F_4,G_4) \nonumber
\end{eqnarray}

Similar tables are easily obtained for $\alpha<0$ and we shall not present them here.
We see that for $\alpha \neq 0$ we must have $k\neq0$ ($k=0$ would imply
$c_1=c_2=0$). Otherwise, k remains arbitrary, as do $\alpha,...,\xi$ in eq.
(\ref{fg}) and (\ref{fg2}). The integration constants $\sigma_1$ and $\sigma_2$
must be such that the polynomials $P_3(f)$ and $Q_3(g)$ in eq. (\ref{fg2}) have
multiple roots. The constants $c_1$ and $c_2$ are completely determined.

Let us sum up our results as a theorem.

\begin{theorem}

Separation of variables in eq. (\ref{intermediaire}) for 
$\ddot{f}^2+g^{''2} \neq 0$ occurs if and only if we have one of the following
\begin{enumerate}

\item
$\alpha=\delta=\xi=0$.\\
In this case we have $g(y)=g_0 y + g_1$ and $f(x)$ is any solution of eq. (\ref{fg})
with $\alpha=0$.

We can put $\phi=-1$,\hspace{0.25cm}$A=g^{'}$ and \hspace{0.25cm}$B=-\dot{f}$.

\item 
$\alpha=0$, $\delta>0$,\\
$g(y)=g_1 e^{\sqrt{\delta} y}+g_2 e^{-\sqrt{\delta} y}-\frac{\xi}{\delta}$, \hspace{1cm}
$g_1 g_2=0$\\
and $f(x)$ is any solution of eq. (\ref{fg}) with $\alpha=0$.

We can put $\phi=-1$,\hspace{0.25cm}$A=g^{'}$ and \hspace{0.25cm}$B=-\dot{f}$.

\item
$\alpha=\beta=\delta=0$, $\xi \gamma < 0$, $c_1=c_2=0$, $k\neq0$, $\phi=\sqrt{-\frac{\xi}{\gamma}}$. \\
In this case we have
$f(x)=\frac{1}{2}\gamma x^2$, \hspace{0.5cm} $g(y)=\frac{1}{2}\xi y^2$\\
and we can put $A= \frac{\phi}{1+\phi}g^{'}$ \hspace{0.25cm} and
\hspace{0.25cm} $B=\phi\dot{f}$.

\item
$\alpha \neq 0$, $k\neq0$, $f(x)$ and $g(y)$ are solutions of eq. (\ref{fg}) as listed
in Table 1 and Table 2, respectively, and $\phi$, $A$ and $B$ as in eq. (\ref{sys3}) where we can put $\tau=\eta=0$. The values of the constants $c_1$ and $c_2$ are listed in the tables
and $\epsilon_1$ and $\epsilon_2$ are related as in eq. (\ref{sol1}) or (\ref{sol2}).

\vspace{0.25cm}

The formulas for the vector potential $(A,B)$ can be modified by putting $A\rightarrow
A + \tau(x)$, $B\rightarrow B + \eta(y)$ without any effect on solutions.

\end{enumerate}
\end{theorem}

\section{Example of solving the Schr\"{o}dinger equation by quasiseparation
of variables}

In order to show how quasiseparation of variables allows us to solve
the Schr\"{o}dinger equation, let us consider the simplest case,
namely (\ref{fg_quad}). We have $f(x)=\frac{\gamma x^2}{2}$, 
$g(y)=\frac{\xi y^2}{2}$ and hence
\begin{eqnarray}
\Omega=\Omega_0=\gamma + \xi, W=\frac{1}{2}(\xi-\gamma+\mu)(\gamma x^2-\xi y^2),
\gamma\xi < 0
\end{eqnarray}

Changing the notation of the constants, we put
\begin{eqnarray}
W=\frac{1}{2}(\omega_1^2x^2+\omega_2^2y^2), \hspace{0.2cm}
\Omega=\Omega_0, \hspace{0.2cm} \omega_1\neq \omega_2, \label{oh}
\end{eqnarray}
In eq. (\ref{sys3}) we choose $\tau(x)=\eta(y)=0$ and obtain
\begin{eqnarray}
A=\frac{\omega_2\Omega_0}{\omega_1+\omega_2}y, \hspace{0.2cm}
B=-\frac{\omega_1\Omega_0}{\omega_1+\omega_2}x, \hspace{0.2cm} \phi(x,y)=\frac{\omega_2-\omega_1}
{\omega_1}.
\end{eqnarray}

In terms of the constant magnetic field $\Omega_0$ and the
frequencies $\omega_1$ and $\omega_2$ the Hamiltonian H and
integral X reduce to
\begin{eqnarray}
H=-\frac{\hbar
^2}{2}(\partial_x^2+\partial_y^2)-i\hbar\frac{\Omega_0}{\omega_1+\omega_2}(\omega_2y\partial_x-\omega_1x\partial_y)
+ \hspace{2.6cm} \label{H_2} \\
\frac{1}{2}\frac{(\omega_1+\omega_2)^2+\Omega_0^2}{(\omega_1+\omega_2)^2}(\omega_1^2x^2+\omega_2^2y^2),
\hspace{1cm} \nonumber
\\
X=-\frac{\hbar^2}{2}\partial_x^2-i\hbar\frac{\Omega_0\omega_1}{\omega_1^2-\omega_2^2}(\omega_2y\partial_x
-\omega_1x\partial_y)+\frac{1}{2}\omega_1^2\frac{(\omega_1+\omega_2)^2+\Omega_0^2}{(\omega_1+\omega_2)^2}x^2.
\end{eqnarray}

The linear combination (\ref{intermediaire}) of H and X that allows the
separation of variables is in this case
\begin{multline}
\{-\frac{\hbar^2}{2}[\frac{\omega_2}{\omega_1}\partial_x^2+\partial_y^2]+\frac{1}{2(\omega_1+\omega_2)^2}[(\omega_1+\omega_2)^2+\Omega_0^2](\omega_1\omega_2x^2+\omega_2^2y^2)\}\psi=\\
(E-\frac{\omega_1-\omega_2}{\omega_1}\lambda)\psi. \hspace{4cm}
\label{s_aux}
\end{multline}
Putting
\begin{eqnarray}
\psi(x,y)=v(x)w(y),
\end{eqnarray}
as in eq. (\ref{separated}) we obtain
\begin{eqnarray}
-\frac{\hbar^2}{2}\frac{\omega_2}{\omega_1}v_{xx}+\frac{1}{2}\frac{\omega_1\omega_2}{(\omega_1+\omega_2)^2}[(\omega_1+\omega_2)^2+\Omega_0^2]x^2v+\frac{\omega_1-\omega_2}{\omega_1}\lambda
v=k_0v, \label{sol_sep}
\\
-\frac{\hbar^2}{2}w_{yy}+\frac{\omega_2^2}{2(\omega_1+\omega_2)^2}[(\omega_1+\omega_2)^2+\Omega_0^2]y^2w-Ew=-k_0w, \nonumber
\end{eqnarray}
where $k_0$ is a separation constant and (E,$\lambda$) are fixed
constants.  The solutions of eq. (\ref{sol_sep}) are of course well
known and the regular, square integrable solutions are given in
terms of Hermite polynomials $H_n$ as
\begin{eqnarray}
v(x)=e^{-\frac{\tau_1^2x^2}{2}}H_{n_1}(\tau_1x), \hspace{0.2cm}
w(y)=e^{-\frac{\tau_2^2y^2}{2}}H_{n_2}(\tau_2y).
\end{eqnarray}
The constants satisfy
\begin{eqnarray}
\tau_a=\sqrt{\frac{\omega_a}{\hbar(\omega_1+\omega_2)}}[(\omega_1+\omega_2)^2+\Omega_0^2]^{1/4},
\hspace{0.2cm} a=1,2,
\end{eqnarray}
\begin{eqnarray}
\nonumber
k_0\omega_1-(\omega_1-\omega_2)\lambda&=&\frac{\hbar\omega_1\omega_2}{2(\omega_1+\omega_2)}\sqrt{(\omega_1+\omega_2)^2+\Omega_0^2}(2n_1+1),\\
E-k_0&=&\frac{\hbar\omega_2}{2(\omega_1+\omega_2)}\sqrt{(\omega_1+\omega_2)^2+\Omega_0^2}(2n_2+1),
\end{eqnarray}
$n_i$=0,1,2,...

Eliminating $k_0$ we have
\begin{eqnarray}
\nonumber E\omega_1
+\lambda(\omega_2-\omega_1)&=&\frac{\hbar\omega_1\omega_2}{\omega_1+\omega_2}\sqrt{(\omega_1+\omega_2)^2+\Omega_0^2}(n+1),
\\
n&=&n_1+n_2. \label{energy}
\end{eqnarray}
The solution $\psi$(x,y) thus depends on two nonnegative integers
$n_1$ and $n_2$, whereas the constants E and $\lambda$ depend only
on n.

Let us now fix E and $\lambda$, i.e. fix $n=n_1+n_2$ and write the
solution of the Schr\"{o}dinger equation with Hamiltonian
(\ref{H_2}) as a superposition of solutions of eq. (\ref{s_aux}):
\begin{eqnarray}
\psi _{E,\lambda}(x,y)=e^{-\frac{1}{2}(\tau _1^2x^2 + \tau
_2^2y^2)}\sum_{n_1=0}^{n}A_{n_1,n-n_1}H_{n_1}(\tau
_1x)H_{n-n_1}(\tau _2y), \label{sol_superposee}
\end{eqnarray}
where the constants $A_{n_1,n-n_1}$ are to be determined.  We
substitute (\ref{sol_superposee}) into the Schr\"{o}dinger
equation, use the recursion relations for the Hermite polynomials
and obtain a linear homogeneous equation for the constants
$A_{n_1,n-n_1}$.  This equation is best written in matrix form
involving a tridiagonal matrix
\begin{eqnarray}
M|A>=|0>,
\end{eqnarray}
i.e.
\begin{eqnarray}
\left(\begin{array}{ccccc} \alpha_{11} & -S&0 & &\\
nS&\alpha_{22}&-2S&\ddots&\\
0&\ddots&\ddots&\ddots&0\\
&\ddots &2S&\alpha_{nn}&-nS\\
& &0&S&\alpha_{n+1n+1}
\end{array}\right)
\left(\begin{array}{c} A_{0n}\\
A_{1n-1}\\  \vdots\\ A_{n-11} \\A_{n0}
\end{array}\right)
= \left(\begin{array}{c} 0 \\ \vdots \\ \vdots \\ \vdots
\\ 0
\end{array}\right)
\label{sys_mat}
\end{eqnarray}

\begin{eqnarray}
\nonumber \alpha_{11}=R(\omega_1+(2n+1)\omega_2)-E , \hspace{0.2cm} \alpha_{22}= R(3\omega_1+(2n-1)\omega_2)-E,\\
\alpha_{nn}= R((2n-1)\omega_1+3\omega_2)-E, \hspace{0.2cm}
\alpha_{n+1n+1}=R((2n+1)\omega_1+\omega_2)-E,
\\
 \nonumber R=\frac{\hbar\sqrt{(\omega_1+\omega_2)^2+\Omega_0^2}}{2(\omega_1+\omega_2)},
\hspace{0.2cm}
S=\frac{i\hbar\Omega_0}{\omega_1+\omega_2}\sqrt{\omega_1\omega_2}.
\end{eqnarray}
The energy is obtained from an algebraic equation of order n+1,
namely
\begin{eqnarray}
detM=0.
\end{eqnarray}
The constant $\lambda$, for E and n given, is obtained from eq.
(\ref{energy}).  The wave function for E and $\lambda$ fixed has
the form (\ref{sol_superposee}) with coefficients $A_{n_1n-n_1}$
obtained by solving the system (\ref{sys_mat}).

Thus, once eq. (\ref{s_aux}) is solved by separation of variables,
the solution of the Schr\"{o}dinger equation reduces to linear
algebra.  The energy operator is reduced to block diagonal form,
with each block of finite dimension, namely n+1.  In this scheme
the problem is "exactly solvable".

For low values of n this can be done explicitly.  For the ground
state we have:
\begin{eqnarray}
 n=0 \hspace{0.5cm}
E&=&\frac{\hbar}{2}\sqrt{(\omega_1+\omega_2)^2+\Omega_0^2},
\hspace{0.2cm}
\lambda=\frac{\hbar\omega_1}{2(\omega_1+\omega_2)}\sqrt{(\omega_1+\omega_2)^2+\Omega_0^2},
\hspace{1cm}
\\
\psi_{00}&=&A_{00}e^{-\frac{\tau_1^2x^2+\tau_2^2y^2}{2}}. \nonumber
\end{eqnarray}
The first two excited states satisfy
\begin{eqnarray}
\nonumber n=1 \hspace{0.5cm}
E_{\pm}=\frac{\hbar}{2}[2\sqrt{(\omega_1+\omega_2)^2+\Omega_0^2}\pm
\sqrt{(\omega_1-\omega_2)^2+\Omega_0^2}], \hspace{3.5cm}\\
\lambda_{\pm}=\frac{\hbar}{2(\omega_2^2-\omega_1^2)}[-2\omega_1^2\sqrt{(\omega_1+\omega_2)^2+\Omega_0^2}
\hspace{4.9cm}
\\ \nonumber \mp(\omega_1+\omega_2)\sqrt{(\omega_1-\omega_2)^2+\Omega_0^2}],  \hspace{2.5cm}\\
\nonumber
\psi_{\pm}=e^{-\frac{\tau_1^2x^2+\tau_2^2y^2}{2}}[A_{01}^{\pm}H_1(\tau_2y)+A_{10}^{\pm}H_1(\tau_1x)],
\hspace{4.2cm}
\end{eqnarray}
with
\begin{eqnarray}
A_{10}^{\pm}=-\frac{i}{2\Omega_0}[\sqrt{(\omega_1+\omega_2)^2+\Omega_0^2}(\omega_1+3\omega_2)\pm
\sqrt{(\omega_1-\omega_2)^2+\Omega_0^2}
(\omega_1+\omega_2)]A_{01}^{\pm}.
\end{eqnarray}
For n=2 and n=3 we must solve a cubic and quartic equation,
respectively.

\section{Conclusions}

 We have shown that in the presence of a magnetic field $\Omega
\neq 0$ the existence of a first order integral of motion leads to
the separation of variables in the Schr\"{o}dinger equation, once
we make the proper choice of coordinates and gauge.

The existence of a second order integral does no longer imply the separation of
variables. In many cases, identified in Theorem 1 of Section 4, we can
"`quasiseparate" variables, that is separate variables in one equation
that is an appropriate linear combination of the two equations,
$H \psi = E \psi$ and $X \psi = \lambda \psi$. The solutions of the
Schr\"{o}dinger equation are then linear combinations of solutions of
this equation, namely (\ref{intermediaire}).

In Section 5 we analysed one of the separable cases in detail, namely
that of a constant magnetic field $\Omega_0$ and an anisotropic harmonic
oscillator effective potential W (see eq. (\ref{oh})).

Thus quasiseparation of variables is reminescent of the Dirac equation
\cite{Kalnins2}. The same method can be used for all cases identified
in Section 4.

The question that remains is the following: What can we do in those
integrable cases when neither separation of variables, nor quasiseparation
of variables occurs? The same questions arise in the case of polar
integrability where the
magnetic fields and effective potentials are known \cite{Berube,McSween},
but the Schr\"{o}dinger equation remains to be solved.  Some examples of
parabolic and elliptic integrability are known \cite{Pucacco}; the solutions
of the Schr\"{o}dinger equation remain to be studied.

Finally, a few words about superintegrability \cite{Evans,Evans2,Fris,
Gravel,Gravel2,Kalnins,Kalnins2,Kalnins4,Makarov,McSween,Sheftel,Tempesta,
Winternitz,Wojciechowski}.  In two dimensions,
superintegrability means that two independent operators, $X_1$ and
$X_2$, commuting with H (but not with each other) exist.  The only
known case with a nonzero magnetic field that exists, is that of a
constant magnetic field $\Omega_0$ and a zero effective scalar
potential W=0.  In this case there are three first order integrals and
they generate a four-dimensional Lie algebra, isomorphic to a
central extension of the Euclidean Lie algebra e(2) \cite{Berube}.

\section*{Acknowledgements}

The research of P.W. was partly supported by a research grant from
NSERC of Canada.  F.C. and C.H. benefitted from NSERC undergraduate
student research awards.

\newpage

\Large{Table 1}\normalsize

Functions f(x) leading to quasiseparation of variables. We have \begin{eqnarray}\nonumber
\omega_1^2=\frac{\sqrt{\beta^2-4\alpha\gamma}}{4}, \beta^2-4\alpha\gamma > 0,
\alpha > 0, f_3-f_1= \frac{6\omega_1^2}{\alpha}, N_1=\frac{8\epsilon_2 N_2}{k\epsilon_1}
\end{eqnarray}

\begin{tabular}{|r|r|r|r|r|}
\hline
No.&$f(x)$&$c_1$&$N_2$&Comment\\
\hline
$F_1$&$\frac{6}{\alpha(x-x_0)^2}-\frac{\beta}{2\alpha}$&$2kf_1$&$\epsilon_2\frac{k}{2}\sqrt{\frac{-k\alpha}{3}}$&$f_1=f_2=f_3\leq f(x)$\\

&&&&$f_1=-\frac{\beta}{2\alpha}$\\
\hline
$F_2$&$f_1+\frac{f_3-f_1}{\sin^2{\omega_1 (x-x_0)}}$&$2kf_3$&$\epsilon_2\frac{k}{2}\sqrt{\frac{-k\alpha}{3}}$&$f_1=f_2<f_3\leq f(x)$\\

&&&&$f_1=\frac{-\beta-\sqrt{\beta^2-4\alpha\gamma}}{2\alpha}$\\
\hline
$F_3$&$f_1+(f_3-f_1){\tanh^2{\omega_1 (x-x_0)}}$&$2kf_1$&$-\epsilon_2\frac{k}{2}\sqrt{\frac{-k\alpha}{3}}$&$f_1\leq f(x) \leq f_2=f_3$\\

&&&&$f_3=\frac{-\beta+\sqrt{\beta^2-4\alpha\gamma}}{2\alpha}$\\
\hline
$F_4$&$f_1+\frac{f_3-f_1}{\tanh^2{\omega_1 (x-x_0)}}$&$2kf_1$&$\epsilon_2\frac{k}{2}\sqrt{\frac{-k\alpha}{3}}$&$f_1<f_2=f_3\leq f(x)$\\

&&&&$f_3=\frac{-\beta+\sqrt{\beta^2-4\alpha\gamma}}{2\alpha}$\\

\hline
\end{tabular}

\vspace{1cm}

\Large{Table 2}\normalsize

Functions g(y) leading to quasiseparation of variables. We have \begin{eqnarray}\nonumber
\omega_2^2=\frac{\sqrt{\delta^2+4\alpha\xi}}{4}, \delta^2+4\alpha\xi > 0,
\alpha > 0, g_3-g_1= \frac{6\omega_2^2}{\alpha}, N_1=\frac{8\epsilon_2 N_2}{k\epsilon_1}
\end{eqnarray}

\begin{tabular}{|r|r|r|r|r|}
\hline
No.&$g(y)$&$c_2$&$N_2$&Comment\\
\hline
$G_1$&$\frac{\delta}{2\alpha}-\frac{6}{\alpha(y-y_0)^2}$&$-2kg_3$&$-\epsilon_1\frac{k}{2}\sqrt{\frac{-k\alpha}{3}}$&$g(y)\leq g_1=g_2=g_3$\\

&&&&$g_3=\frac{\delta}{2\alpha}$\\

\hline
$G_2$&$g_3-\frac{g_3-g_1}{\tanh^2{\omega_2 (y-y_0)}}$&$-2kg_3$&$-\epsilon_1\frac{k}{2}\sqrt{\frac{-k\alpha}{3}}$&$g(y)\leq g_1=g_2<g_3$\\

&&&&$g_1=\frac{\delta-\sqrt{\delta^2+4\alpha\xi}}{2\alpha}$\\
\hline
$G_3$&$g_3-(g_3-g_1){\tanh^2{\omega_2 (y-y_0)}}$&$-2kg_3$&$\epsilon_1\frac{k}{2}\sqrt{\frac{-k\alpha}{3}}$&$g_1=g_2\leq g(y) \leq g_3$\\

&&&&$g_1=\frac{\delta-\sqrt{\delta^2+4\alpha\xi}}{2\alpha}$\\
\hline
$G_4$&$g_3-\frac{g_3-g_1}{\sin^2{\omega_2 (y-y_0)}}$&$-2kg_1$&$-\epsilon_1\frac{k}{2}\sqrt{\frac{-k\alpha}{3}}$&$g(y)\leq g_1<g_2=g_3$\\

&&&&$g_3=\frac{\delta+\sqrt{\delta^2+4\alpha\xi}}{2\alpha}$\\

\hline
\end{tabular}

\end{document}